\documentclass[prc,twocolumn,showpacs]{revtex4}

\usepackage{graphicx}%
\graphicspath{{./PaperFigs/}{./}}
\usepackage{dcolumn}
\usepackage{amsmath}
\usepackage{amssymb}

\renewcommand{\vec}[1]{\boldsymbol{#1}}
\newcommand{\ph}{\vec{\phi}}

\newlength\graphwidth
\begin{document}
 
\title{Soliton Systems at Finite Temperatures and Finite Densities}

\author{Oliver Schwindt\email{schwindt@yahoo.de} and Niels R. Walet\email{Niels.Walet@umist.ac.uk}}
\affiliation{Dept.\ of  Physics, UMIST, Manchester M60 1QD, UK}
\date{\today}

\begin{abstract}
The finite-density and finite-temperature phase portraits of the
baby-Skyrme and Skyrme models are investigated. Both grand-canonical
and canonical approaches are employed. The grand-canonical approach
can be used to find the ``natural'' crystal structure of the
baby-Skyrme model and it is shown to have triangular symmetry. The
phase portraits include solid, liquid and phase-coexistence between
solids and vacuum states. Furthermore, a chiral phase transition can
be observed for both models.  The phase portrait of the Skyrme model
is compared with that expected of strongly
interacting matter, and we also contrast our results to other
models.\end{abstract}


\maketitle

\section{Introduction}\label{Introduction}

One of the great questions in the theory of strong interactions is the
behavior of baryonic matter at finite density and finite
temperatures. Even though it is widely believed that quantum
chromo-dynamics (QCD) is the correct theory to describe the strong
interactions, its non-perturbative nature makes it extremely difficult
to describe the low energy consequences of the theory.  Lattice QCD
\cite{mu:quantum} is the preferred numerical technique to study the
QCD Lagrangian directly. Here, QCD is modeled non-perturbatively by
discretising the theory on a lattice, and a numerical approximation is
made to the path integral describing either the Euclidean time
evolution or the thermodynamics of the theory.  The masses of mesons,
baryons, and glueballs have already been calculated to reasonable
accuracy \cite{gu:introduction}. The limitations of the approach are
both numerical (momentum cut-off, finite lattice size), and more
theoretical.  The most important difficulty is the description of
fermions on the lattice.  Since lattice QCD relies on using Monte
Carlo methods, computing power is another of the key limitations. If
sufficient powerful computers were available, the inaccuracies due to
discretising QCD could be reduced by adopting larger
gridsizes. Therefore, as computing power increases in future, more
accurate results will be achieved.  Even then, dealing with a finite
density of fermions is an especially tricky problem, since the
fermionic determinant to be calculated becomes complex, leading to
various instabilities.

Due to these difficulties, it is quite common to use models to study
the phase diagram of strongly interacting matter. Recent examples of
such approaches can for instance be found in studies by Rajagopal {\it
et al}~\cite{ra:mapping,st:signatures}, who describe the phase diagram
using a Nambu-Jona-Lasino model~\cite{be:color} and a random matrix
model~\cite{ha:on}.  The QCD phase diagram, obtained from a model with
two massless quarks in Ref.~\cite{ra:mapping}.  
A typical phase diagram shows a hadronic-matter phase, the quark-gluon plasma, and a
color superconducting phase. Earth's surface, in general, has a low
temperature (0.025\,eV) and a chemical potential much less than
1\,GeV, and therefore we live in a (liquid) hadronic-matter phase. The
most striking feature of the phase diagram is the presence of a
critical point. This separates an area
where a first-order phase transition occurs, from one 
where
a continuous path can be taken from the hadronic phase to the
quark-gluon plasma phase. At the critical point itself, it is believed
that there is a second-order phase transition. The position of the
critical point is extremely difficult to model and has been estimated
to have a temperature $T_{E}\approx 140-190$\,MeV and a chemical
potential of $\mu_E\approx
200-800$\,MeV~\cite{ra:mapping,ra:traversing}.  The color superconducting
phase cannot be modeled by the Skyrme model because color-flavor
locking is observed, and in order to model it, three colors need to
be considered rather than the $N_c=\infty$ approximation underlying
the Skyrme model, which will be discussed later. The color superconducting
phase will therefore not be discussed further.

The Skyrme model has a venerable history \cite{sk:model} in the
non-perturbative description of nucleon structure and the low-energy
behavior of baryonic matter, since it contains a good description of
the long-wave length behavior of the dynamics of hadrons. As has been
argued by 't Hooft and Witten \cite{ho:planar,ho:two,wi:current}, this
is closely related to the large-number-of-colors limit of QCD, in
which baryons must emerge as solitons, in much the same way as happens
in the Skyrme model. Alternatively, we can interpret the model in
terms of chiral perturbation theory, in terms what is much like a
gradient expansion in terms of the pion field.  The model can be used
to describe, with due care \cite{br:deuteron,br:novel}, systems of a
few nucleons, and has also been applied to nuclear and quark matter.
Within the standard zero-temperature Skyrme model description there
are signatures of chiral symmetry restoration at finite density, but
in a rather special way, where a crystal of nucleons turns into a
crystal of half nucleons at finite density, which is chirally
symmetric only on average \cite{ja:phase}. The question of the
finite-temperature has never been addressed and would be of some
interest.  We shall also look at the two-dimensional Skyrme model,
which has the advantage of being easier to interpret, but also has
physical relevance; a special form of the two-dimensional Skyrme model
has recently been developed for use in quantum Hall systems
\cite{Girvin}. This model is obtained as an effective theory when the
excitations relative to the $\nu=1$ ferromagnetic quantum Hall state
are described in terms of (a gradient expansion in) the spin density,
a field with properties analogous to the pion field in nuclear physics
\cite{LK90}. Apart from obvious changes due to the number of
dimensions, the new approach differs from the historical Skyrme model
by having a different time-dependent term in the Lagrangian, and the
appearance of a non-local interaction, where the topological charge
density at different points interacts through the Coulomb force. In
the limit of large Skyrmions this last term can be approximated by a
more traditional local ``Skyrme term'', which is quartic in the
fields, leading to the standard baby-Skyrme model with local
interactions.  Even with the non-local complication the model is on
the whole remarkably similar to the nuclear Skyrme model. The
Skyrme-field effective degrees of freedom describe the ground state of
such systems, and probably also the low energy dynamics and
thermodynamics, so that we can ask similar questions about the
finite-density physics as for baryonic matter.

In this paper we shall consider the local baby Skyrme model, which can
be regarded as the 2D version of the traditional 3D nuclear one. We
can also extrapolate to one dimension. The model obtained in that way
is the sine Gordon model expressed in terms of an unit complex
field. All of these models contain a topologically conserved charge,
which in the three-dimensional case is identified with baryon
number. This gives rise to topological solitons, which in nuclear
physics are identified as baryons. Since the behavior of nuclear
matter at high densities is expected to reveal, probably by one or two
phase transitions, the substructure of baryonic matter, it is
interesting to study this phenomenon in Skyrme models. At the same
time it allows us to look at the phase diagram of quantum Hall
Skyrmionic systems, which may well be the easiest way to calculate
parts of the phase diagram of the underlying electronic model.

At finite density but zero temperature the Skyrme models and the sine
Gordon model have a crystalline structure, which consist of regularly
spaced solitons. In the one-dimensional case, where exact solutions
exist \cite{gu:investigation}, this is known to be a zero temperature
artifact, and we have a liquid at any finite temperature, no matter
how small. In the nuclear Skyrme model we would like to find a fluid,
to mimic the quantum liquid behavior expected in nuclear matter
\cite{co:mean}.  In order to appreciate the subtleties involved, one
must understand that the Skyrme model, as a classical field theory,
can be understood as a semiclassical (large action) limit of a quantum
theory.  Clearly the quantum fluctuations in the underlying theory
could be large enough to wash out the crystalline structure, as happens
for the sine Gordon model. In the special case of one space dimension
it is also well known that thermal and quantum fluctuations play
exactly the same role, and thermal fluctuations also break the
crystalline state. For the Skyrme and baby Skyrme models it is not easy
to access the quantum fluctuations, since the field theories are
non-renormalisable, but we can access the thermal fluctuations (with
care, because nonremormalisability plays a role in the thermodynamics
as well!).  Furthermore, we might well wish to study the physics of
these systems at temperatures where thermal fluctuations dominate the
physics. Therefore, we shall concentrate on the finite temperature
phases of these classical field theories.

The most direct approach to the problem is to perform a Monte-Carlo
(Metropolis algorithm based) study of the partition function of each
of these models.  As we shall argue below the least obvious aspect in
such an approach is how to deal with the topological conservation
laws. Rather than immediately tackle the 3D model, we shall first
concentrate on the (local) 2D model. This has the advantage that the
visualization of, and also the understanding gained by, the results is
much more straightforward.  The extensions to quantum-Hall Skyrmions
are under way and will be presented elsewhere
\cite{we:QHE}. We expect the current for the QHE solitons to be very 
similar to the results reported here.

This paper is organized as follows. In Sec.~\ref{sec:formalism} we
discuss the formalism succinctly. In the next section,
Sec.~\ref{sec:results} we discuss in detail the results obtained in
our simulations (a small excerpt has already been published in
Refs.~\cite{os:phase,os:lett}). We draw some conclusions, and give an
outlook for possible future work in Sec.~\ref{sec:conclusions}.

\section{Formalism}\label{sec:formalism}

\subsection{Lagrange density}

The Lagrange density for Skyrme models can be given as~\cite{sk:model,sk:unified}
\begin{eqnarray}
 {\cal L}&=&\frac{1}{2}(\partial_{\mu}\phi_{k})^{2}-
\frac{1}{4}(\partial_{\mu}\phi_{k}\,\partial^{\mu}\phi_{k}\,\partial_{\nu}\phi_{l}\,\partial^{\nu}\phi_{l})
\nonumber\\&&+
\frac{1}{4}(\partial_{\mu}\phi_{k}\,\partial^{\mu}\phi_{l}\,\partial_{\nu}\phi_{k}\,\partial^{\nu}\phi_{l})+
\frac{m_{\pi}^{2}}{f_{\pi}^{2}e^{2}}(\phi_0-1)\label{Lagphi}~,
\end{eqnarray}
which as long at we take the space index $\mu=0,1,\ldots,d$ to
describe $d+1$ dimensional space, and the field components
$k,l=1,\ldots,d+1$ to describe a $d+1$ dimensional unit vector, can
describe the Sine-Gordon theory ($d=1$), the baby Skyrme model ($d=2$)
and the nuclear Skyrme model ($d=3$).  The Lagrange density can be
separated into a kinetic and a potential term
\begin{equation}
 {\cal{L}}={\cal T}[\phi]-{\cal V}[\phi]~,
\end{equation}
where 
\begin{eqnarray}
 {\cal T}[\phi] &=& \frac{1}{2}\partial_{t}\phi_{a}\,\partial_{t}\phi_{b}\,{\cal M}_{ab}[\phi]~,\\
 {\cal M}_{ab}[\phi] &=& \delta_{ab}+\partial_{i}\phi_{c}\,\partial_{i}\phi_{c}\,\delta_{ab}-\partial_{i}\phi_{a}\,\partial_{i}\phi_{b}~,\label{massmatrix}\\
 {\cal V}[\phi] &=& \frac{1}{2}\partial_{i}\phi_{a}\,\partial_{i}\phi_{a}+\frac{m_{\pi}^{2}}{f_{\pi}^{2}e^{2}}(1-\phi_0)
\nonumber\\&&+
\frac{1}{4}([\partial_{i}\phi_{a}\,\partial_{i}\phi_{a}]^{2}-[\partial_{i}\phi_{a}\,\partial_{j}\phi_{a}][\partial_{i}\phi_{b}\,\partial_{j}\phi_{b}])
~.\label{potentialV}
\end{eqnarray}

For our later calculation we need we need the expression for the
energy density, ${\cal T}+{\cal V}$, which for the Skyrme model
becomes
\begin{equation}
 {\cal E} = \frac{1}{2}\partial_{t}\phi_{a}\,\partial_{t}\phi_{b}\,{\cal M}_{ab}(\ph(\vec{x}))+{\cal V}(\ph(\vec{x}))~.\label{energydensity}
\end{equation}
The two sets of constraints (implemented by the delta-functions in
the integrals) are $\ph^2-1=0$ and $\dot{\ph}\cdot\ph=0$ at each point
in space, where the first constraint forces~$\ph$ to have unit length
and the second constraint follows by taking the time derivative of the
first constraint.

In 3D the topological charge density, which was identified by Skyrme
with the baryon-number density operator, is given by
\begin{eqnarray}
 {\cal B}(\vec{x})&=&
\frac{1}{12\pi^{2}}\varepsilon^{0\sigma\rho\nu}\varepsilon^{\alpha\beta\gamma\delta}\phi_{\alpha}\,\partial_{\nu}\phi_{\beta}\,\partial_{\sigma}\phi_{\gamma}\,\partial_{\rho}\phi_{\delta}\nonumber\\
&=&\frac{1}{2\pi^{2}}\det(\ph,\partial_{x}\ph,\partial_{y}\ph,\partial_{z}\ph)~,\label{baryondensity}
\end{eqnarray}
with simpler expressions in 1D and 2D.
The energy density~(\ref{energydensity}) and baryon-number
density~(\ref{baryondensity}) are then inserted into the
grand-canonical partition function~(\ref{derivedpartfunc}).

\subsection{Partition Function}
\label{ThePartitionFunction}

In order to study the thermodynamics of a theory, we must construct a
partition function. We shall discuss both grand-canonical and
canonical partition functions, which are also discussed in 
Refs.~\cite{os:thesis,ha:simulated}. The explicit expressions for the
thermodynamic partition functions will be given for the baby-Skyrme
and the Skyrme models. The Metropolis principle is used to evaluate
the partition functions, see Refs.~\cite{os:thesis,ha:simulated}
for more details.

\subsubsection{The Grand-Canonical Partition Function}\label{GrandCanonPartFunc}

The general thermodynamic partition function for a general unitary  field
theory on the lattice is given by
\begin{eqnarray}
{\cal Z}&=&\int\prod_{p=1}^{M^d}d^{d+1}\phi_{p}\,d^{d+1}\dot{\phi}_{p}\,\delta(\dot{\ph}_{p}\cdot\ph_{p})\,\delta(\ph^2_p-1)
\nonumber\\&&\qquad\qquad\times\exp(-\beta(E-\mu B))~,\label{derivedpartfunc}
\end{eqnarray}
where an integration is to be performed over the field and field
derivative at each lattice point. The quantities $E$ and $B$ are the
total energy and baryon number of the system.

If the field makes small vibrations (which must be harmonic) at each
lattice point then each degree of freedom contributes $\frac{1}{2}k_B
T$ to the free energy. There are $M^d$ lattice sites, each with $d$
potential and $d$ kinetic degrees of freedom. The total energy due to
lattice vibrations is thus $E_{\text{vib}}=\frac{2dM^d}{2\beta}$. The
energy due only to solitons, which we shall refer to as the scaled
energy, is
\begin{equation}
 E_{\text{scaled}}=E_{\text{total}}-E_{\text{vib}}=E_{\text{total}}-\frac{dM^d}{\beta}~.
\end{equation}
It is clear that these vibrational contributions are unphysical and need to be
removed, since in the continuum limit their contribution would
become infinite. The sine-Gordon model is renormalisable, and by
subtracting the energy due to the lattice vibrations, we obtain the
contribution due entirely to the solitons in a similar way as was done
analytically in Ref.~\cite{gu:investigation}. The baby-Skyrme and
Skyrme models are non-renormalisable, and we find that it is therefore
not possible to remove all the lattice-dependency from the energy
expression. However, by removing the harmonic contribution, the
lattice-dependency is reduced. We will not use the energy of a system
to examine phase transitions, because of these
lattice-dependencies. Instead, we use other, less lattice-dependent
methods, which we will discuss in Secs.~\ref{TDBaby} and
\ref{TDFull}.

It is useful to use density-like quantities rather than total
quantities, and therefore we often use quantities like the average
potential-energy density ${\cal V}=\frac{V}{h^d M^d}$ and the average
baryon density ${\cal B}=\frac{B}{h^d M^d}$, where $h^d M^d$ is the
volume of the system. Results are even more comparable when measuring
$\frac{{\cal V}}{{\cal B}}$ (or $\frac{V}{B}$), which gives the
potential energy per soliton. By measuring density-like quantities, we
are measuring quantities which remain finite and meaningful in the
limit of an infinite simulation volume.

The chemical potential $\mu$ determines the particle density in a
particular system. We initially expected that a simulation of the
grand-canonical partition function automatically includes the
probability of adding or removing a soliton from the
system. Unfortunately, because we are working with extended particles,
a soliton can only be added or removed from the system if a number of
field vectors on our discretised lattice change simultaneously. The
probability of this happening is extremely small, and has never been
observed in our simulations. The only examples of the total particle
number changing is if the numerics become unstable. In this case,
neighboring vectors are pointing in quite different directions such
that the derivatives, and therefore the energy and baryon densities,
are not calculated correctly. The numerical breakdown, i.e.\ the
unphysical loss of winding number, is dependent on the lattice
spacing, because it occurs at lower densities and temperatures if the
lattice spacing is increased. Such numerical breakdown must occur
when we have less than the  minimum number of lattice points that can
describe a Skyrmion.

To solve the problem that the number of particles in a given
simulation is fixed by the topology, we use an open system, where
particles can flow through the boundaries freely and therefore we have
a heatbath that acts as a source of particles. The method we use to
implement our open system will be discussed in
Secs.~\ref{TDBaby}, and~\ref{TDFull}, when we discuss the
thermodynamics of our soliton models.

\subsubsection{The Canonical Partition Function}\label{CanonPartFunc}

The canonical approach is usually applied to a closed system, i.e.\
where particles are trapped in a box, and the number of particles in
this box remain constant. As already mentioned in the last section,
the number of solitons is conserved by the topology of the
system. Therefore, the canonical approach seems to be very suitable to
investigate the thermodynamics of soliton systems. The average density
of particles is therefore defined by the initial choice of the number
of particles $B$ in the volume $V$ and it never changes throughout the
simulation. We choose not to implement fixed boundaries, which have
the vacuum value on the simulation edges, but instead we use periodic
boundaries. A system with periodic boundaries can be interpreted as an
approximation of an infinite system because the simulation box is
effectively duplicated an infinite number of times. Periodic boundary
conditions imply that if a Skyrmion flows through one boundary of the
simulation, then it reappears through the opposite side. Periodic
boundary conditions have two advantages over fixed boundaries; firstly
that the simulation can be interpreted as an infinite system, and
secondly that we avoid unphysical boundary effects occurring from
Skyrmions interacting with the fixed boundaries.

The general expression for the canonical partition function is
\begin{eqnarray}
 {\cal Z}_B&=&\int\prod_{p=1}^{M^n}d^{n+1}\phi_{p}\,d^{n+1}\dot{\phi}_{p}\,\delta(\dot{\ph}_{p}\cdot\ph_{p})\,\delta(\ph^2_p-1)
\nonumber\\&&\qquad\qquad\times\exp(-\beta E)~,
\end{eqnarray}
where all the symbols are again defined as in
Sec.~\ref{GrandCanonPartFunc}. The harmonic contributions to the
potential energy are removed in a similar manner as for the
grand-canonical approach,
\begin{equation}
V_{\text{scaled}}=V_{\text{total}}-\frac{nM^n}{2\beta}~.
\end{equation}
In Ref.~\cite{ha:simulated} the zero-temperature minimal-energy
solutions solutions are calculated using simulated annealing within
the canonical approach. For such static solutions, a renormalisation
is irrelevant. Furthermore, the kinetic energy is irrelevant since the
total energy is equal to the potential energy.


We now use the fact that $\dot{\ph}$ is an eigenvector of the mass
matrix (\ref{massmatrix}) to perform the integral over $\dot{\ph}$
analytically. The partition function now simplifies to
\begin{align}
 {\cal Z}={\cal N}^{-1}\beta^{-\frac{d M^d}{2}}
\int \prod_{p=1}^{M^d}d^{d+1}\phi_{p}
&\left(\sqrt{\frac{1+\partial_{i}\phi_{c,p}\,\partial_{i}\phi_{c,p}}{\det({\cal M}_{p})}}\,\right)\nonumber\\
&\times \exp\left(-\beta(V_{p}-\mu B_{p})\right)\label{fullpart}~,
\end{align}
where $V_p$ and $B_p$ are the potential energy and baryon number at
the lattice site $p$ and represent the potential energy and baryon
number in the surrounding unit cell. The overall
constant factor ${\cal N}^{-1}$ is irrelevant when applying the
Metropolis principle, and will be ignored from now
on. Eq.~(\ref{fullpart}) will therefore be used as the grand-canonical
partition function. The lattice does not need to be cubic, but we have
adopted the convention that all the sides have equal length in order to
avoid the additional notation required when using unequal lengths.

The procedure discussed above can be applied to the canonical
formalism, and as expected we obtain the expression (\ref{fullpart})
with $\mu=0$.


The partition function~(\ref{fullpart}) 
cannot be evaluated analytically, and therefore a Metropolis
algorithm is applied, see Refs.~\cite{os:thesis,ha:simulated} for more
details.

The only exception is the sine-Gordon model.  The grand-canonical
partition function is derived in Ref.~\cite{gu:investigation}.
It can very easily be compared to our simulations, since it is obtained
by discretising the field and then taking the continuum limit. Like
many other field-theory models, the observables tend to infinity as
the continuum limit is being reached.  It thus illustrates clearly
the role of renormalisation.

The canonical partition function of the sine-Gordon model in the angle
representation is given by
\begin{widetext}
\begin{eqnarray}
{\cal Z}_B &=&\int \prod_{p=1}^{M}d^{2}\alpha_{p}\,d^{2}\dot{\alpha}_{p}\,\exp\left(-\beta(\frac{1}{2}\partial_{t}\alpha_{p}\,\partial_{t}\alpha_{p}+\frac{1}{2}\partial_{x}\alpha_{p}\,\partial_{x}\alpha_{p}+1-\cos\alpha_{p})h\right)\nonumber\\
  &=& {\rm const}\sqrt{\frac{1}{\beta}}^M\int \prod_{p}d^{2}\alpha_{p}\,\exp\left(-\beta(\frac{1}{2}\partial_{x}\alpha_{p}\,\partial_{x}\alpha_{p}+1-\cos\alpha_{p})h\right)~,
\end{eqnarray}
\end{widetext}
where $\alpha$ is the field, and $\sqrt{\frac{1}{\beta}}^M$ originates
from evaluating the time derivative analytically, see
Sec.~\ref{GrandCanonPartFunc}. The lattice spacing $h$ is required to
convert the energy density to the energy in a cell from the discrete
lattice. The constant factor in the expression for ${\cal Z}_B$ is
irrelevant when measuring thermodynamic quantities, and is therefore
set equal to one. The reason every lattice point behaves like a
classical harmonic oscillator is that the kinetic energy and potential
energy are quadratic in time derivatives and spatial derivatives,
respectively. The contribution to the fluctuations at individual
lattice points from the potential $1-\cos(\alpha)$ is negligible. The
energy at a single lattice point due to fluctuations in the field is
given by $\frac{(\alpha_{i+1}-\alpha_{i})^2}{2h^2}\times h$, or
$\frac{(\alpha_{i+1}-\alpha_{i})^2}{2h}$, where $i$ is the lattice
index and $h$ is the lattice spacing. Therefore,
\begin{align}
{\cal Z}_B &= \left(\frac{2\pi}{\beta}\right)^{M/2}&&\int
\prod_{i=1}^{M}\exp\left(-\beta\sum_{i}\frac{(\alpha_{i+1}-\alpha_{i})^2}{2h}\right)
\nonumber\\&&&
\qquad \times d\alpha_{1}\ldots d\alpha_{M}\nonumber\\
&\approx \left(\frac{2\pi}{\beta}\right)^{M/2}&&\left(\frac{{\rm
const}\times h}{\beta}\right)^{M/2}~. 
\end{align}
Using $P=\frac{1}{\beta L}\frac{\partial {\cal Z}_B}{\partial\beta}$,
the harmonic oscillator contribution to the pressure becomes
$P=\frac{M}{L\beta}\ln(\frac{{\rm const}\times h}{\beta})$. Since the
harmonic oscillator at every lattice point gives a finite contribution
to the pressure $P$, it becomes infinite in the continuum limit. The
constant is not evaluated explicitly here because its contribution to
the energy which we are interested in is irrelevant. Furthermore, the
lattice spacing $h$ within the logarithm also vanishes. The energy
becoming infinite in the continuum limit can be overcome by removing
the contribution from the harmonic oscillators. This way, all
quantities are calculated without contributions from the harmonic
oscillators, and contain only contributions that are not divergent in
the thermodynamic limit. Therefore, the pressure due entirely to the
solitons is given by
\begin{equation}
 P'=-\left(\frac{A\left[-\frac{i\mu\beta}{\pi},4\beta^2\right]}{8\beta^2}+1\right)~.
\end{equation}

We find that the density and internal-energy density for a given
$\beta$ and $\mu$, which are obtained by using the standard
thermodynamic relationships~\cite{ma:statistical,ev:statistical}
$\rho=\left(\frac{\partial P}{\partial\mu}\right)_{\beta}$ and
$u=-\left(\frac{\partial (P\beta)}{\partial\beta}\right)_{\mu\beta}$
respectively, are
\begin{equation}
 \rho=-\frac{1}{8\beta^2}\frac{\partial A\left[-\frac{i\mu\beta}{\pi},4\beta^2\right]}{\partial\mu}\label{MathieuRho}
\end{equation}
and
\begin{equation}
 u=\left(\frac{\partial\left(\frac{1}{8\beta}A\left[-\frac{i\mu\beta}{\pi},4\beta^2\right]+\beta\right)}{\partial\beta}\right)_{\mu\beta}~.\label{densitySG}
\end{equation}

\section{Results}\label{sec:results}
\subsection{Sine-Gordon model}
\begin{figure}[!tbp]
\includegraphics[width=\graphwidth]{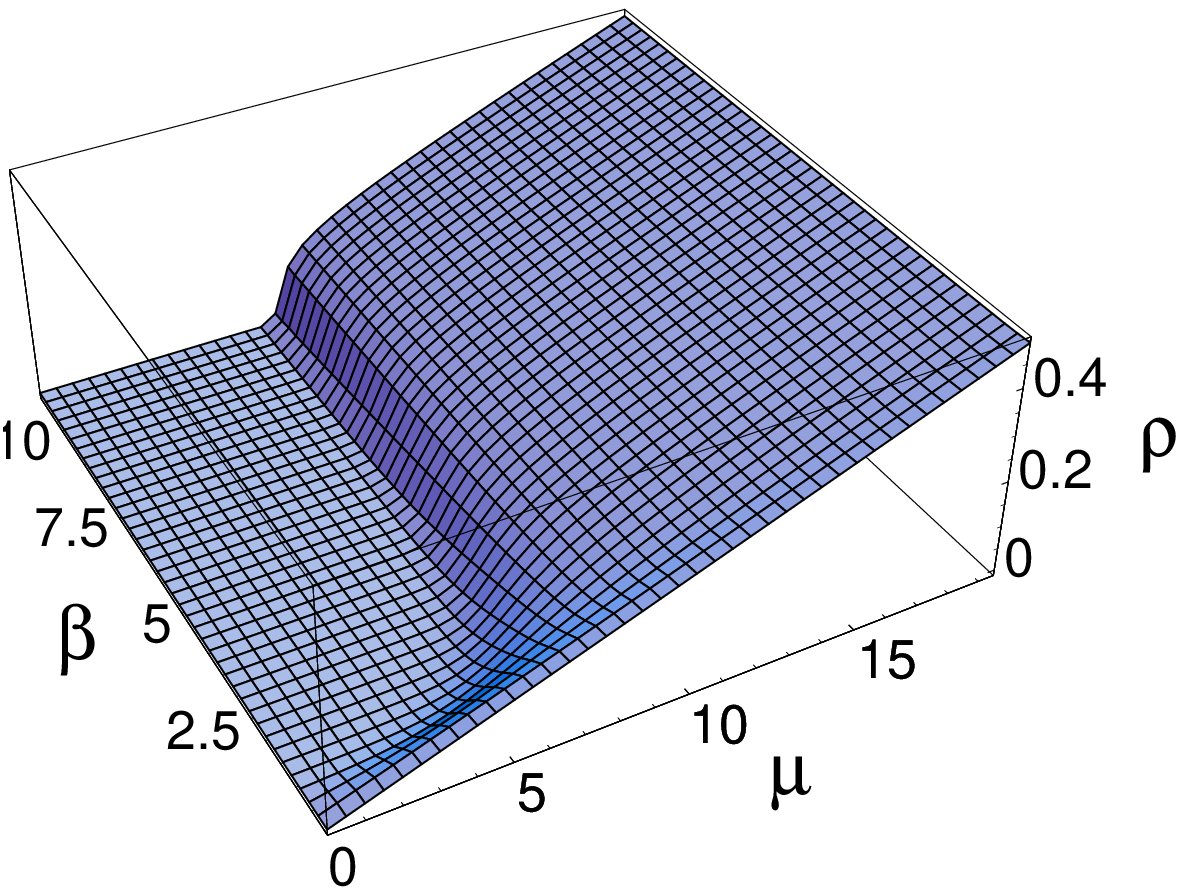}\\
\includegraphics[width=\graphwidth]{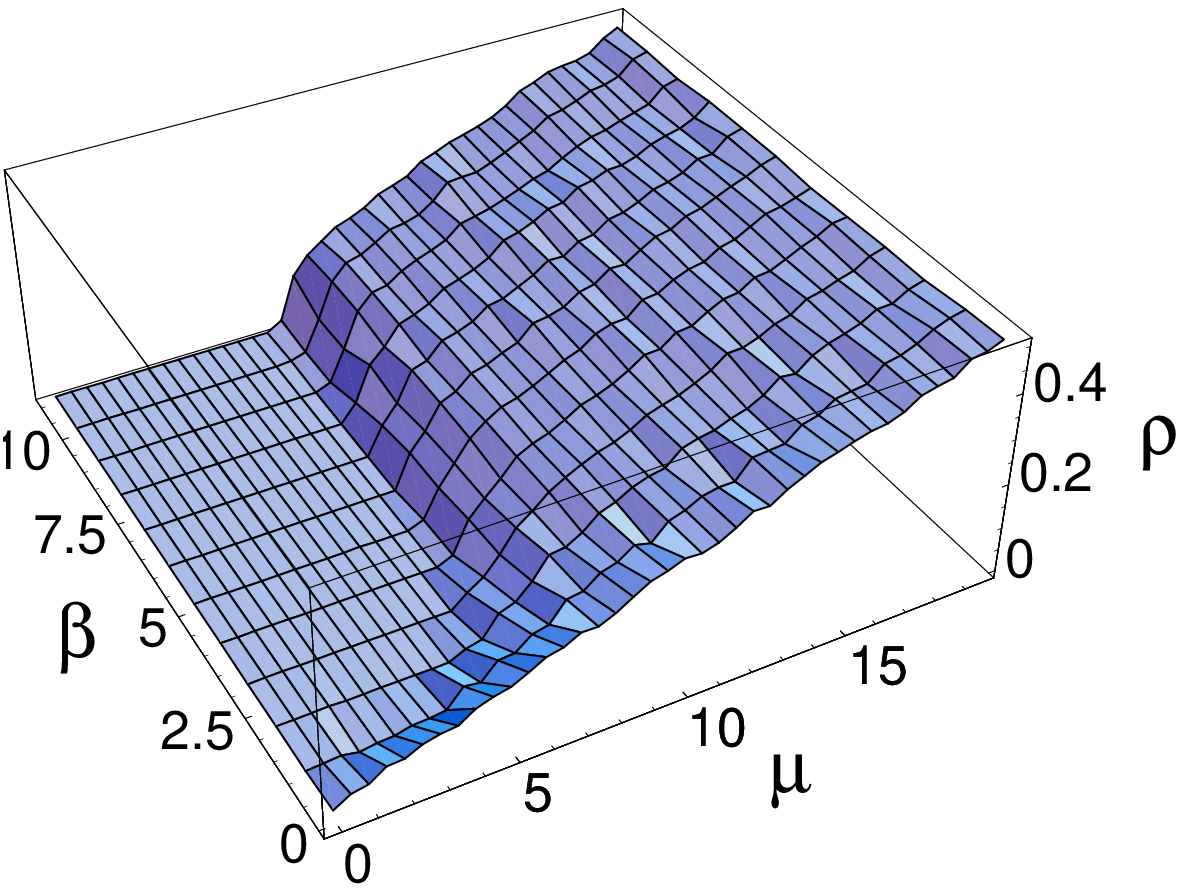}
\caption{The exact (top figure) an numerical (lower figure)
results for the grand-canonical equation of state of the sine-Gordon model.}\label{fig:MuBetaRho}
\end{figure}

A check of the calculations can be made  by comparing the exact results as
discussed in the previous section to a calculation using open
boundary conditions. This means that we impose no boundary conditions
at all on the finite simulation volume, but allow solitons
to enter the system through those boundaries. Even though, in principle,
one can create topological charge on a grid, in practice
topological charge is very well conserved. The only way to
perform a   grand-canonical simulation in practice is thus using
open boundary conditions. We then compare
the simulated result for the density of solitons to
the analytic result, Eq.~(\ref{MathieuRho}), 
in 
Fig.~\ref{fig:MuBetaRho} as a function of $\beta$ and $\mu$.
As we can see  the results are 
very similar,with some statistical noise in the simulated results.
Since it is much more straightforward to perform large simulations
for the sine-Gordon model than for the 2D and 3D Skyrme models, this
shows the best quality of results that can be expected (apart from
a limitation in simulation time, since we have looked at quite a dense
set of grid for the results).
\subsection{Thermodynamics of the Baby-Skyrme Model}\label{TDBaby}

\subsubsection{The Grand-Canonical Approach}\label{babygrand}


The special feature of the grand-canonical approach is that we use
open-boundary conditions. This means that the number of particles
present in a particular simulation can change when particles enter or
leave the system over the boundary. The lattice points at a boundary
behave slightly differently than the internal ones.  If a field vector
is sampled at a boundary, it is accepted or rejected depending only on
the change in the integrand on the neighboring internal
plaquettes (we use the word plaquette for a square enclosed
by four nearest neighbor lattice points). The field vectors at the boundaries are therefore less
restricted than those at other lattice points.



\begin{figure}[!tbp]
\includegraphics[width=\graphwidth]{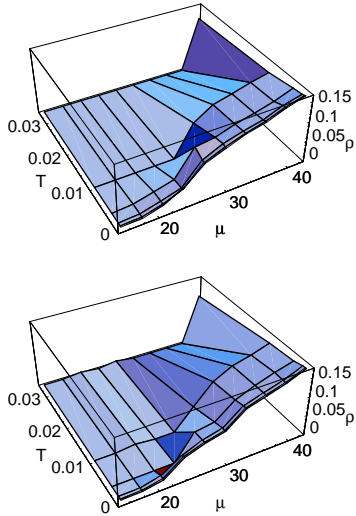}
\caption{The $\mu,~\rho,~T$ phase portrait for the baby-Skyrme
model. {\it Top:}~The order in which the points in the plot are
calculated is in the increasing $T$ direction with constant
$\mu$. {\it Bottom:}~The calculation order is in the decreasing $T$
direction with constant $\mu$. The plots differ because the initial
field configuration at each point is the final configuration of the
previously calculated point. The axes are labeled in Skyrme units.}
\label{BTDparam}
\end{figure}

\begin{figure}[!tbp]
\includegraphics[width=\graphwidth]{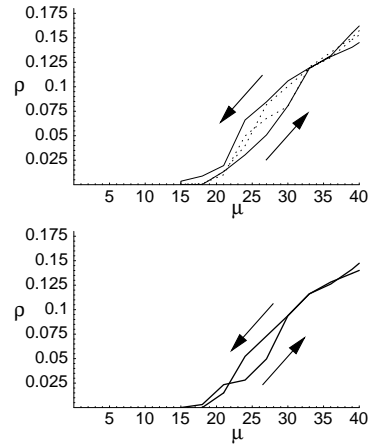}
\caption{The hysteresis effect when increasing and then decreasing the
chemical potential~$\mu$. {\it Top:}~The solid lines, where the
arrows mark the direction of the simulation, show the hysteresis
effect for a low temperature $T=0.0001$. The dashed lines are the
results imported from Fig.~\ref{BTDparam}, where the temperature was
changed for constant~$\mu$. {\it Bottom:}~For higher temperatures, the
hysteresis effect is less visible. The axes are labeled in Skyrme
units.}
\label{BTDHyst}
\end{figure}

The $\mu,~\rho,~T$ phase diagram for the baby-Skyrme model shown in
Fig.~\ref{BTDparam}. Each point is a separate simulation which has
reached equilibrium for a given chemical potential~$\mu$ and a
temperature~$T$. To save computing time, the initial field
configuration used to calculate the density~$\rho$ at each point in
the plot was the final configuration for the previously calculated
temperature. The arrows in Fig.~\ref{BTDparam} show the order the
simulations where calculated in. Two plots of the same phase portrait
have been shown where the order of calculation has been reversed. The
reason for showing this is because the computing time required to
reach equilibrium becomes large near the phase transition, and the
simulation time was not long enough to reach equilibrium near the
phase transition. By comparing the two graphs, one can see that it is
difficult to predict the chemical potential of the transition to
within 5~Skyrme-energy units per soliton. In Fig.~\ref{BTDHyst}, the
net density of baby-Skyrmions~$\rho$ is shown against the chemical
potential~$\mu$ at a fixed temperature~$T$, for two different
temperatures. In each plot, the order in which $\mu$ changes has been
shown; one in the increasing $\mu$ direction and one in the decreasing
$\mu$ direction. For low temperatures, the hysteresis effect is more
visible than for higher temperatures, because solitons do not move as
quickly and hence it takes longer to reach equilibrium. The magnitude
of the hysteresis effect depends on the computing time used, as
equilibrium is reached extremely slowly and we did not have the time
to let the system strictly reach it.

Each baby-Skyrmion has to have an energy greater than the topological
lower bound for it to exist in an open system
simulation. Fig.~\ref{BTDHyst} shows that the chemical potential,
which defines how much energy is given to each soliton, must be
greater than approximately 18~Skyrme-energy units per baby-Skyrmion,
$\mu\gtrsim 18$. Once the chemical potential is greater than the
threshold, the density increases approximately linear with the
chemical potential, $\rho\propto\mu-\mu_{\text{threshold}}$.

As long as $\mu$ is large enough, which is the case for almost all of
the finite-$\rho$ region, the state of the system is crystalline. The
open boundaries allow crystal structures to form without defects, and
therefore we can determine the ``natural'' crystal structure which
will be described in the next section. The liquid state and the phase
coexistence between solid and vacuum states exist near the phase
transition and are too difficult to model using the grand-canonical
approach. These states will be examined using the canonical approach
in Sec.~\ref{BabyCanonical}.


\begin{figure}
\includegraphics[width=\graphwidth]{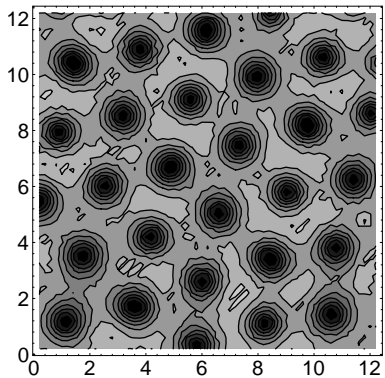}
\caption{Baryon-number density plots of the natural crystal structure
of the baby-Skyrme model with open-boundary conditions. The
baryon-number density is averaged over 20000 steps. The axes are labeled in
Skyrme-length units. Lighter shades represent higher baryon-number
densities. The parameters for this simulation are $\mu=30$, $T=0.05$,
and the number of solitons in the simulation volume is about 13 ($rho=xxx$).}
\label{BabyNatPlot}
\end{figure}

For high densities and low temperatures, the Skyrmions merge into a
lattice where it is not possible to identify individual Skyrmions, see
Fig.~\ref{BabyNatPlot}. Since there are no fixed boundaries in this
simulation, there are no defects in the crystalline solutions
either. The lattice is determined most easily from the structure of
the ``holes'' with zero baryon-number density. There are two such
minima present per baryon. In this crystal phase, the average field
$\langle\ph\rangle$ is zero, and therefore a mean-field chiral
symmetry exists.

The triangular nature of the crystal is evidence that the type of
crystal favored by the baby-Skyrme model is independent of the
lattice used to discretise the model. Even though the lattice is
square, the crystal which forms does not align itself to it. Since
this simulation allows the crystal structure having the lowest energy
per baryon to form without being effected by boundary conditions, we
call it the ``natural'' crystal structure. The field configuration
within a unit cell was copied (since it satisfies periodic boundary
conditions) and simulated annealing with periodic boundary conditions
was used to accurately investigate the zero-temperature properties.

\begin{figure}[!tbp]
\includegraphics[width=\graphwidth]{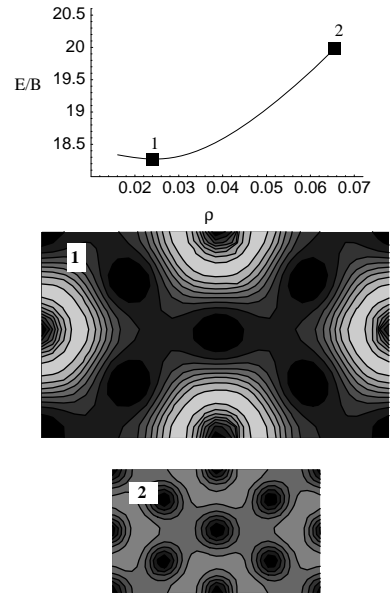}
\caption{The high-density phase at zero temperature. {\it Top:}~Energy
per baryon against density. {\it Middle:}~A baryon-number density plot
of the lowest energy per baryon number attainable, i.e.\ the
``natural'' crystal structure. Lighter shades represent higher
baryon-number density. {\it Bottom:}~A baryon-number density plot at
high density. The baryon number is more evenly distributed throughout
the structure. The middle and right plot are drawn to the same scale.}
\label{Tri}
\end{figure}

The correct high-density structure~\cite{os:phase} where Skyrmions
merge into a lattice is shown in Fig.~\ref{Tri}. The lattice is
determined most easily from the structure of the ``holes'' with zero
baryon-number density, since it is not possible to identify individual
Skyrmions. There are two such minima present per baryon. In the
low-density phase, the average field $\langle\phi_k\rangle$ is
polarized in the direction of the sigma field, but for high densities,
the average field is zero. Therefore, a mean-field chiral symmetry
exists in the high-density phase and not in the low-density phase.
The energy per baryon ratio when minimizing the energy of a field
configuration having the form of the natural crystal is
18.27~Skyrme-energy units at a density of 0.024 particles per cubic
Skyrme-length unit. If the density is lowered further, then the
minimal-energy solution consist of small numbers of Skyrmions bound
together, i.e.\ the multi-Skyrmions. In fact, this can be defined as
the point where the phase transition between the high-density and
low-density matter occurs.

\subsubsection{The Canonical Approach}\label{BabyCanonical}


As we have argued before, with periodic boundary conditions it seems
impossible to generate additional topological charge. Nevertheless
these calculations are much preferable over the open boundary
calculations, since the results suffer much less from artifacts due to
those boundaries (but the results will now depend much more strongly
on the symmetries of the initial seed to the simulation). This is
still the preferred way to calculate the phase diagram.  In order to
create the $\rho,~T,~\frac{u}{\rho}$ phase diagram, the
lattice-dependent terms must be removed from the internal
energy. Unfortunately, the lattice-dependency cannot be removed
entirely, because the baby-Skyrme model is non-renormalisable.

The internal energy per unit area $u$ is found by using the
thermodynamic relationship
\begin{equation}
 u=U/L^2=-\frac{1}{L^2}\left(\frac{\partial\ln {\cal Z}}{\partial\beta}\right)_{B}~,
\end{equation}
where $L^2$ is the simulation volume and $B$ denotes the constant
particle number defined by the initial conditions. Therefore,
\begin{widetext}
\begin{eqnarray}
 u&=&\frac{M^2}{L^2\beta}+\frac{1}{L^2}\frac{\int \prod_{p}d^{3}\phi_{p}\,\left(\sqrt{\frac{1+\partial_{i}\phi_{c,p}\,\partial_{i}\phi_{c,p}}{\det({\cal M}_{p})}}\right)V_{p}\exp\left(-\beta(V_{p})\right)}{\int \prod_{p}d^{3}\phi_{p}\,\left(\sqrt{\frac{1+\partial_{i}\phi_{c,p}\,\partial_{i}\phi_{c,p}}{\det({\cal M}_{p})}}\right)\exp\left(-\beta(V_{p})\right)}
= 
 \frac{1}{h^2 \beta}+\frac{1}{L^2}\langle V\rangle~, 
\end{eqnarray}
\end{widetext}
where $h$ is the lattice spacing and $\frac{1}{L^2}\langle V\rangle$
is the potential energy per unit area.

One can attempt to remove some of the lattice-dependency by assuming
that the fluctuations at lattice points behave like harmonic
oscillators, and therefore assuming that
the contribution from the potential energy of the harmonic oscillator,
$\frac{1}{h^2\beta}$, can be removed, so
\begin{equation}
 u_{\text{scaled}}=\frac{1}{L^2}\langle
 V\rangle-\frac{1}{h^2\beta}\nonumber~.
\end{equation}
Unfortunately $u_{\text{scaled}}$ still depends strongly on the
lattice spacing. Furthermore, for high temperatures,
$u_{\text{scaled}}$ can become negative, making its interpretation
difficult. If the number of solitons in the system is zero,
$u_{\text{scaled}}$ still has a non-linear behavior that can become
negative, when we would expect it to be zero. 
This is clearly an indication of the non-renormalisability of the
Skyrme model, and we can no longer use the simple technique (count
the lattice degrees of freedom) used to renormalise the sine-Gordon model

Instead of searching for a detailed theoretical explanation of the
lattice-dependency, which would require the study of counter terms,
we chose to compare simulations with a fixed
number of solitons with simulations that use the same lattice
parameters and are at the same temperature, but contain zero
solitons. Thus, we use
\begin{equation}
 u_{\text{scaled}} =\frac{1}{L^2}\langle V\rangle-\frac{1}{L^2}\langle V_0\rangle\label{Buscaled}~,
\end{equation}
where $\langle V\rangle$ is the potential energy of a simulation with
solitons, and $\langle V_0\rangle$ is the potential energy of the same
simulation without any solitons. Although there is still some
lattice-dependency, it is not very strong. It is likely that it is not
possible to completely remove the dependency on the lattice spacing
because the model is non-renormalisable. Nonetheless, we use this
method because the internal energies for different
lattice spacing become quite comparable, and phenomena
such as phase transition points do not seem to depend on the lattice
scale chosen.

\begin{figure}
\includegraphics[width=\graphwidth]{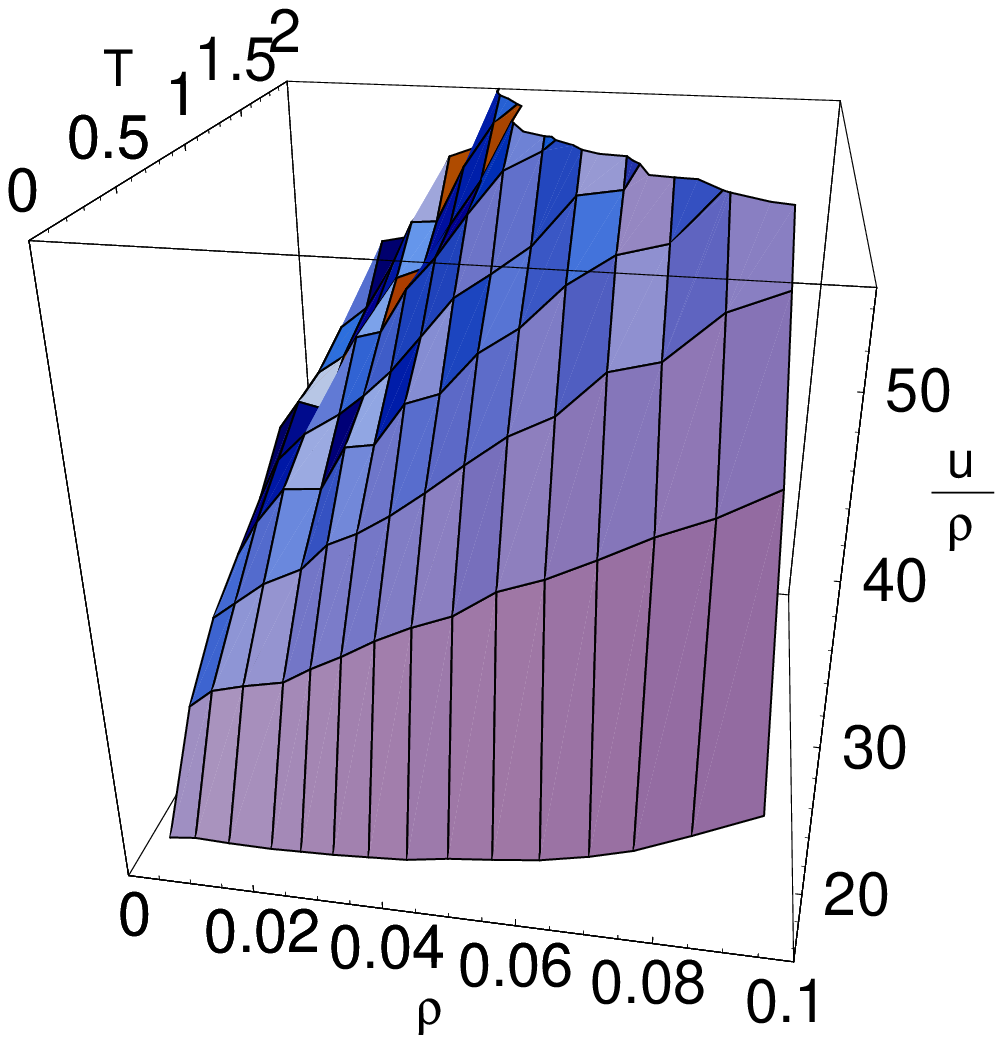}
\caption{The $\rho,~T,~\frac{u}{\rho}$ phase diagram. Some of the
contribution from finite-lattice effects is removed by using
Eq.~(\ref{Buscaled}) to find the internal-energy density. The
density~$\rho$ is altered by changing the lattice spacing and keeping
the same number of Skyrmions in the system. The axes are labeled in
Skyrme units. (16 solitons on a $90\times90$ lattice.)}
\label{BTDRhoBetaEnergy}
\end{figure}

We first use a canonical approach to study the $\rho,~T,~\frac{u}{\rho}$
phase diagram. 
The internal energy per soliton is plotted against the density~$\rho$
and the temperature~$T$ in Fig.~\ref{BTDRhoBetaEnergy}. The internal
energy is measured using Eq.~(\ref{Buscaled}). The density~$\rho$ is
altered by changing the lattice spacing~$h$ while keeping the number
of Skyrmions in the system at a constant value. Although
Eq.~(\ref{Buscaled}) is used to remove some of the lattice-dependency,
there is still a remaining contribution.

At finite temperatures, vibrations at each lattice point are present,
but the contribution to the energy depends on the number of lattice
points per unit volume. Since we vary the lattice spacing to change
the soliton density, the internal energy per soliton as shown in
Fig.~\ref{BTDRhoBetaEnergy} may not be properly compared for different
densities. This phase diagram will not be discussed further, because
it cannot be used to interpret the states of matter throughout the
$\rho$-$T$~plane. In the next two sections, the methods that
successfully determine the states of matter for the baby-Skyrme model
throughout the $\rho-T$~plane are discussed.

\begin{figure}
\includegraphics[width=\graphwidth]{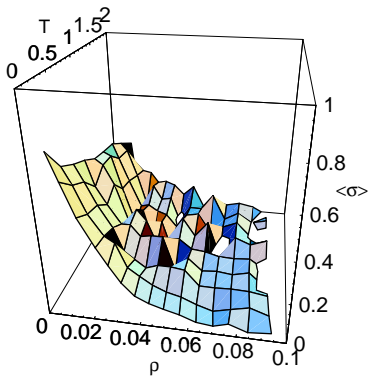}
\caption{The chiral symmetry (magnetization) 
$\langle\sigma\rangle$ as a function of $\rho$ and $T$. For fluids and
solids, we find $\langle\sigma\rangle\approx 0$. For the phase
coexistence between solid and vacuum, $\langle\sigma\rangle> 0$. $T$
and $\rho$ are given in Skyrme units and $\langle\sigma\rangle$ is
dimensionless. The parameter $\langle\sigma\rangle$ has been measured
from a single field configuration and has not been averaged through
time, and the results show large fluctuations. All simulations were
performed for 16 solitons on a $90\times 90$ lattice, and the density
was varied by changing the lattice spacing.}
\label{BTDRhoBetaChiral}
\end{figure}

Along the lines first proposed by Klebanov~\cite{kl:nuclear}, we
analyse the grid-averaged fields $\langle\ph \rangle$. The average of
the pion fields~$\langle\vec{\pi}\rangle$ is always zero, but the
average sigma field~$\langle\sigma\rangle$ changes. The average sigma
field~$\langle\sigma\rangle$ is plotted against the temperature~$T$
and the density~$\rho$ in Fig.~\ref{BTDRhoBetaChiral}.

When few Skyrmions exist in a large volume,
$\langle\sigma\rangle>0$. This is because the vacuum, where
$\sigma=1$, contributes a significant amount to the average sigma
field. For high densities, however, the baby-Skyrmions combine in such
a way that the individual baby-Skyrmions cannot be
identified. Furthermore, an approximately chirally-symmetric
configuration is formed where $\langle\sigma\rangle\approx 0$ and the
sigma field and the pion fields are interchangeable without change in
energy. For the baby-Skyrme model, chiral symmetry is never exactly
satisfied (in a time average), because the pion-mass term, which is
required to stabilize the solitons, violates this symmetry. As the
density of a system is increased, $\langle\sigma\rangle=0$ is
approached asymptotically. In Fig.~\ref{BTDRhoBetaChiral}, a phase
transition in~$\langle\sigma\rangle$ can be observed to occur between
the densities $0.02<\rho<0.04$ for various temperatures.

The fluctuations in~$\langle\sigma\rangle$ are largest in the region
$0.02<\rho<0.08$, $T>0.5$. In the next section, we identify a liquid
and a solid phase in the chirally symmetric phase, and the liquid
region is located where the fluctuations in $\langle\sigma\rangle$ are
largest. The liquid and solid states possibly extend into the broken
chiral symmetry phase, depending on where one defines the chiral phase
transition to be. These fluctuations are present because
in Fig.~\ref{BTDRhoBetaChiral} 
$\langle\sigma\rangle$ was calculated from a single field
configuration and not averaged in time. Also, 
the~$\langle\sigma\rangle$ phase portrait seems to be independent of
the lattice parameters, unlike the internal-energy density. When
doubling the number of lattice point and keeping the same number of
baby-Skyrmions in the same volume, the~$\langle\sigma\rangle$ phase
diagram is not significantly different than the one shown in
Fig.~\ref{BTDRhoBetaChiral}.


By examining the pseudo-time-averaged baryon-density plots at various
temperatures and densities, one can easily identify three different
states that a particular simulation may be in, namely in a solid, in a
liquid, or in a phase-coexistence state. Furthermore, correlation
functions may be used to draw the same conclusions. In this section,
the number of solitons in a system is varied to create different
densities, while the lattice size and the lattice spacing remain
fixed. Although the results shown have the same lattice parameters,
the same conclusions can be drawn in simulations with different
lattice parameters, as has been checked for a number of examples.

\begin{figure}
\includegraphics[width=\graphwidth]{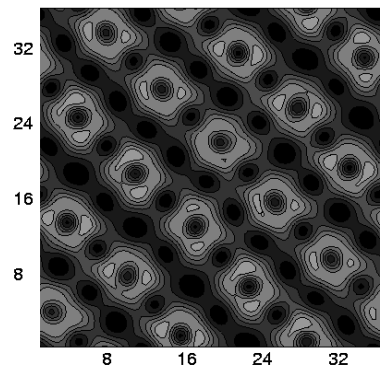}
\caption{A simulation in the solid phase.   
Pseudo-time-averaged baryon-number density plot. Lighter shades represent
higher baryon-number density and the units shown are in Skyrme-length
units. The density $\rho=0.04$, which implies that each light-colored circle
contains two units of topological charge.}
\label{Solid}
\end{figure}

\begin{figure}
\includegraphics[width=\graphwidth]{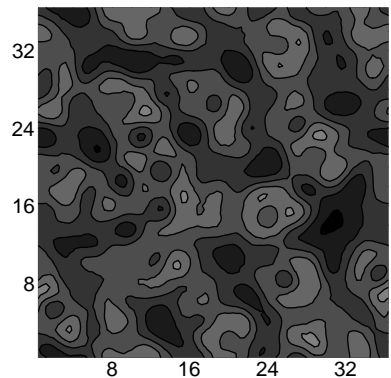}
\caption{A simulation in a liquid phase, as a pseudo-time-averaged
baryon-density plot with an average taken with $2000$ pseudo-time
steps. The shades from dark to light represents low to high
baryon-number densities and the units shown are in Skyrme-length
units. As in Fig.~\ref{Solid} $\rho=0.04$, but the temperature is  higher.}

\label{LiquidBW}
\end{figure}

\begin{figure}
\includegraphics[width=6cm]{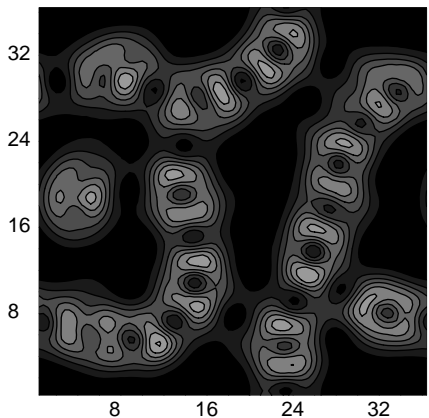}
\caption{Pseudo-time averaged baryon-number density plot in the
phase-coexistence region.  The shades from dark to light represents
low to high baryon-number densities.
As in Figs.~\ref{Solid},\ref{LiquidBW} $\rho=0.04$, but the temperature is 
intermediate between the two.}
\label{Coexist}
\end{figure}

An example of the baryon-number density distribution for a solid is
given in Fig.~\ref{Solid}. On the left we show a pseudo-time-averaged
baryon-number density plot and on the right a snapshot where thermal
fluctuations are evident. The approximate triangular crystal structure
can be observed although defects are present, which are a consequence
of using periodic boundary conditions, in contrast to the absence of
defects seen when using the grand-canonical approach.

A pseudo-time-average of a liquid shows an almost constant
baryon-number density, because structures move, see
Fig.~\ref{LiquidBW}. The visible structure in a pseudo-time-averaged
baryon-density plot is dependent on the number of iterations used to
create it, in this case $2000$ pseudo-time steps.

In the phase-coexistence region, one observes series of multi-solitons
that do not change with pseudo-time, i.e.\ a solid structure
surrounded by large regions of vacuum. These can either be interpreted
as a crystal of nuclei, or as a percolating network.  An example is
shown in Fig.~\ref{Coexist}.

The solid, liquid and phase-coexistence states can also be identified
by examining correlation graphs. The correlation graphs are created by
plotting the correlation between the baryon number at the lattice
points, 
\begin{align}
 C(r)=\sum_{a=1}^{\substack{\text{all~lattice~}\\ \text{points}}}
\sum_{b=1}^{\substack{\text{~all~lattice}\\ \text{points}}}
&\left(\langle B^2_a B^2_b\rangle-\langle B_a^2\rangle\langle B_b^2\rangle\right)
\nonumber\\&\times
\delta(r-|\vec r_a-\vec r_b|)
,
\end{align}
against $r$, which is the separation between lattice points $a$ and
$b$. The correlation function $C$ at $r$ is calculated by taking the
average of the correlation with respect to all $a$ and $b$ separated
by a distance $r$. To keep track of the value $B^2_a B^2_b$ for each
separation distance, we bin the values, and therefore the distances
are rounded to the nearest half-lattice spacing, i.e.\ to the nearest
$h/2$.

\begin{figure}
 \includegraphics[width=5cm]{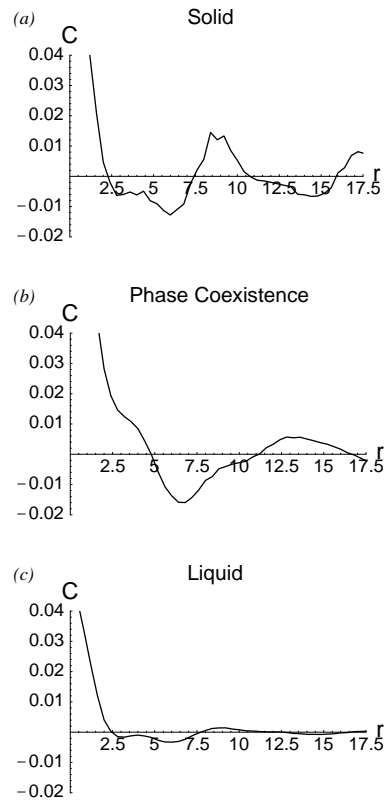} 
\caption{The correlation functions for a solid, a state in phase
coexistence, and a liquid, respectively. {\it (a):}~For the solid,
there are peaks a distance of several Skyrme-length units away from
the origin because the crystal is repetitive. {\it (b):}~For phase
coexistence, there are also peaks a distance of several Skyrme-length
units away from the origin because the baby-Skyrmions are bound
together, but the peaks are weaker than for a simulation in the solid
phase, and lack the fine structure. {\it (c):}~For a liquid, the
correlation function decreases rapidly to zero after a few
Skyrme-length units. The non-zero component exists because of the
correlation of Skyrmions with themselves. ($r$ is in Skyrme-length
units and $C$ is dimensionless.)}
\label{CorrBaby}
\end{figure}

The correlation graphs can be used to identify the state of a
system. For solids and phase coexistence (which is a solid surrounded
by vacuum), correlation functions show peaks a distance of several
Skyrme-length units away from the origin, see Fig.~\ref{CorrBaby}. For
liquids, however, the correlation function levels to zero quickly. The
non-zero component is the correlation of Skyrmions with themselves,
and is therefore non-zero for approximately the length of the radius
of a baby-Skyrmion. Some structure is visible to a distance of
10~Skyrme-length units because interactions with neighboring
Skyrmions are still present. It is because of the visible long-range
interaction that we interpret the state to be a liquid rather than a
gaseous state. The advantage of using correlation functions rather
than pseudo-time-averaged baryon-density plots to identify liquids is
that they are not dependent on the number of iterations used to create
them.

\begin{figure*}[!tbp]
\includegraphics[width=\textwidth]{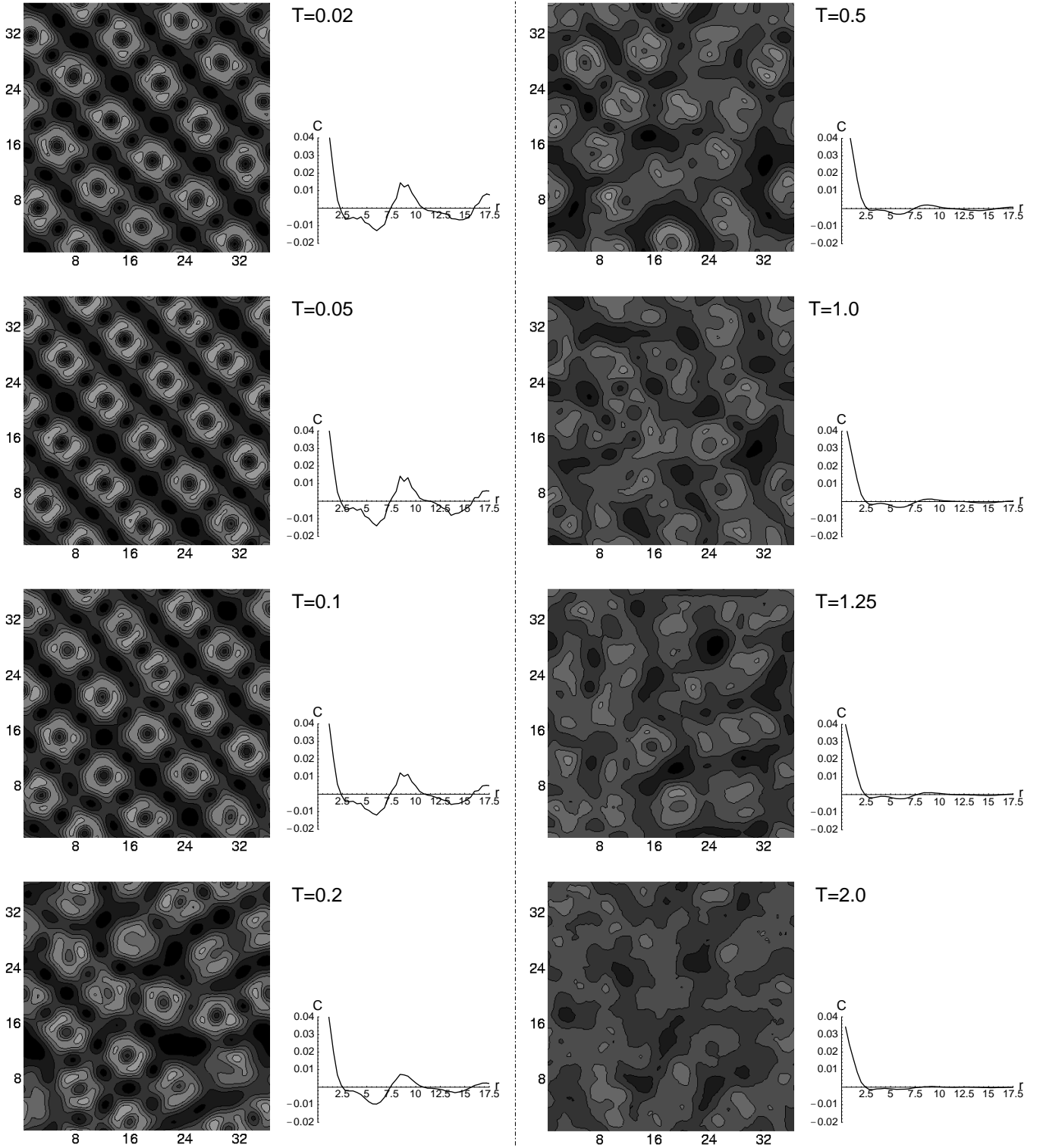}
\caption{Pseudo-time-averaged baryon-density plots and the
corresponding correlation functions for various temperatures with
density $\rho=0.028$. The length units shown are in Skyrme-length
units and the correlation units are dimensionless. The contour levels for
all plots are identical where white illustrates the highest baryon
density. All simulations are performed on a $90\times 90$ size lattice with
lattice spacing $h=0.4$. The simulations whose baryon densities are
plotted on the left have all been identified as solids and those
plotted on the right have been identified as liquids.}
\label{BTDG1}
\end{figure*}

The transition from solid to liquid as the temperature is raised at a
constant density is shown in Fig.~\ref{BTDG1}. Both the
pseudo-time-averaged baryon-density plots and the corresponding
correlation functions are shown to highlight the fact that the state
of the system can be identified through either method. The simulations
whose baryon densities are plotted on the left have all been
identified as solids and those plotted on the right have been
identified as liquids. The pseudo-time average for these plots was set
to $2000$ iterations, and therefore it can be seen that the baryons
move around more for simulations at higher temperature. When using
more iterations, the baryon density approaches the average everywhere,
but because there is less structure, it defeats the purpose of
these illustrations. It is not possible to identify the structure
at $T=0.2$ directly from the baryon-density plot, but the correlation function
clearly shows that it is a solid. The apparent motion of the solitons
are lattice vibrations.

\begin{figure}
\includegraphics[width=\graphwidth]{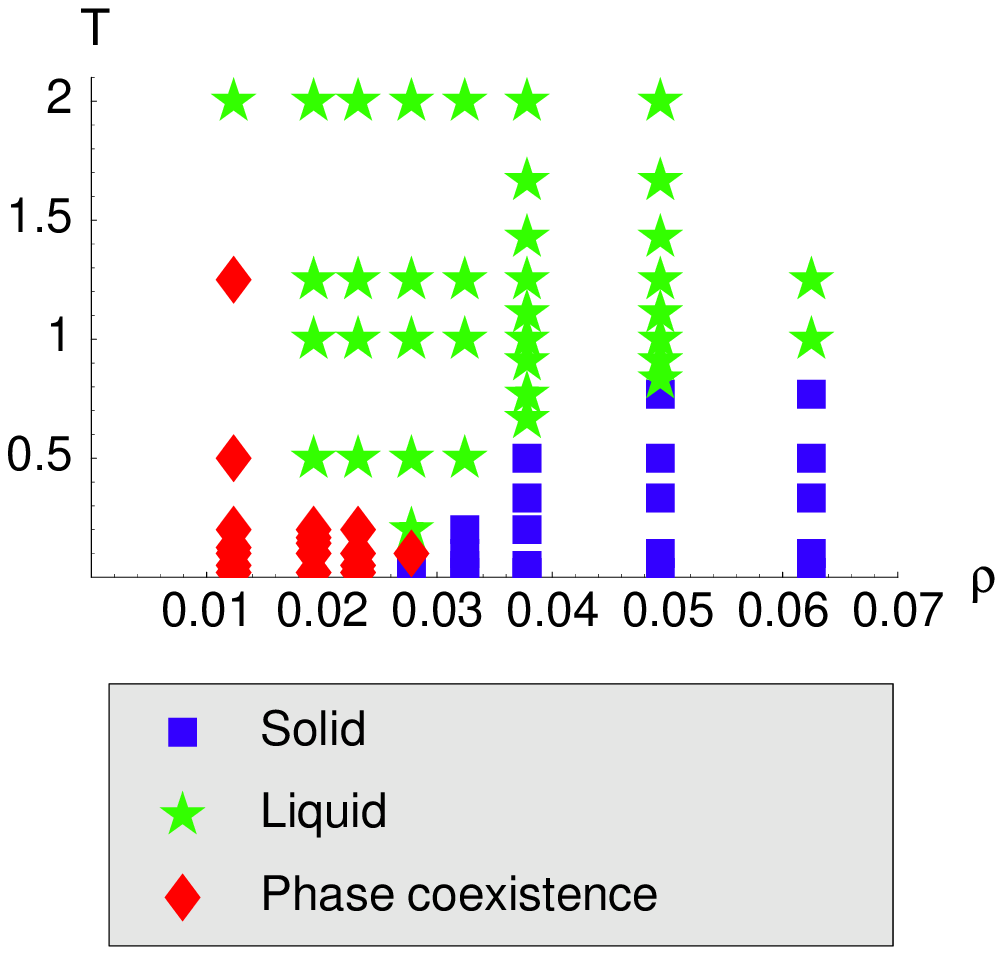}
\caption{The phase diagram. All plotted points are results obtained
from simulations performed on a $90\times90$ lattice with a lattice
spacing of $h=0.4$ Skyrme-length units. The axes are labeled in Skyrme
units.} 
\label{phasediagram}
\end{figure}

Fig.~\ref{phasediagram} shows the final phase portrait for the
baby-Skyrme model which has been created by examining the
pseudo-time-averaged baryon-density plots and correlation functions at
various temperatures and densities. We identify solid, liquid, and
phase-coexistence regions, by combining the correlation function measure
with that of the time-averaged baryon density.

Unfortunately, we are unable to determine the order of any of the
phase transitions. It is unlikely that any of them are of first order,
because we would expect to see discontinuities in the
$\rho,~T,~\frac{u}{\rho}$ and $\rho,~T,~\langle\sigma\rangle$ phase
diagrams. Since higher-order phase transitions can only be identified by
discontinuities in the $n$-th order derivative,
we are unable to do so since , the
results presented here contain statistical fluctuations that are large
enough such that such an analysis becomes too difficult.

\subsubsection{Conclusions}

The baby-Skyrme model shows a rich phase diagram. By examining the
chiral symmetry, the time-averaged baryon-density plots, the snapshot
baryon-density plots and the correlation functions, it was possible to
identify the state of matter that exists for a given density and
temperature. Although the model is non-renormalisable and the internal
energies are dependent on the lattice parameters, all the other
measured quantities used seem to be unaffected. Therefore, we are
confident that the mapping of the phase portrait of the baby-Skyrme
model is given to a reasonable accuracy.

A similar diagram will be created for the 3D Skyrme model, and the
phase diagrams are expected to differ because the stabilizing terms
have different forms. In particular, we would like to know at which
temperatures and densities the phase transitions occur and compare
these with putative QCD phase diagram such as those in
Ref.~\cite{ra:mapping}.

\subsection{Thermodynamics of the Skyrme Model}\label{TDFull}

In the previous section, we have successfully investigated the phase
portrait of the baby-Skyrme model. We observed solid states, liquid
states, and phase coexistence between solid and vacuum states. The
internal energy, even after removing some of the lattice contribution,
is still dependent on the lattice spacing because the baby-Skyrme
model is non-renormalisable. However, by examining the chiral symmetry
breaking, correlation functions, and baryon-density plots, we were
able to determine the state of the system with methods that do not
depend on lattice spacing. We will repeat the same type of analysis
for the Skyrme model. Since the Skyrme model is often argued to be
an approximation to
QCD, the Skyrme phase portrait will be compared to other model of QCD 
(specifically Ref.~\cite{ra:mapping}, which was created using another
model). The phase transition between nuclear matter and
``quark'' matter is of particular current interest.

Unfortunately, we were unable to apply the grand-canonical approach to
the Skyrme model. If the initial condition is a vacuum throughout the
simulation box, we do not observe solitons entering the
system. Skyrmions are expected to enter the system if $\mu>65$
Skyrme-energy units per soliton. Simulations for a number of
choices of $\mu$, $T$ and $h$ suffered from numerical
breakdown, which was identified by neighboring field vectors pointing
in almost opposite directions, where all derivative
approximations underlying the lattice approach break down.
 Numerical breakdown seems to occur more easily for the
Skyrme model than for the baby-Skyrme model, most likely because the
model is more complex. In the cases where a numerical breakdown did
not occur, the initial vacuum never changed. Either solitons could not
flow over the boundaries, or we did not have the computational time
required to observe any solitons entering the system.

The field configuration that gives the lowest energy per soliton that
has yet been observed is given by Castillejo {\it
et~al}~\cite{ca:dense} and the energy per soliton is $3.8$\% above the
topological lower bound. It is not clear whether this simple-cubic
lattice in half-Skyrmions is the natural crystal structure. The
periodic boundary conditions used in that reference favor this
configuration. In fact, since we see that the baby-Skyrme model has a
triangular pattern, we would expect that a hexagonal close-packed
structure is a candidate for the natural crystal structure. We were
not able to create a canonical simulation where we had the correct
number of particles in a simulation box that was shaped to favor a
hexagonal close-packed crystal. This does not imply that it is not the
natural crystal structure, but instead it means that guessing the
correct environment without knowing exactly what this crystal looks
like is extremely difficult. The natural crystal structure cannot be
identified from the canonical simulations either, because large
simulation volumes are required, and we do not have the computational
resources to do this for the Skyrme model. Because of the large
simulation volume required to identify the natural crystal structure
from a canonical simulation, we probably would not have discovered the
triangular crystal for the baby-Skyrme model without using the
grand-canonical approach.

\subsubsection{The Canonical Approach}

The canonical approach for the nuclear-Skyrme model is implemented in
a similar manner as for the baby-Skyrme model. 
Again, some of the lattice-dependency is removed from the internal
energy. The internal energy of the solitons is given by
\begin{equation}
 u_{\text{scaled}}=\frac{1}{L^3}\langle V\rangle-\frac{1}{L^3}\langle V_0\rangle~,\label{FullScaled}
\end{equation}
where $\langle V\rangle$ is the average potential energy of a
simulation with finite density and inverse temperature $\beta$. The
average potential energy $\langle V_0\rangle$ is calculated from a
simulation with identical parameters but without any Skyrmions, i.e.\
it is the potential energy due entirely to vibrations on the lattice.

\begin{figure}
\includegraphics[width=\graphwidth]{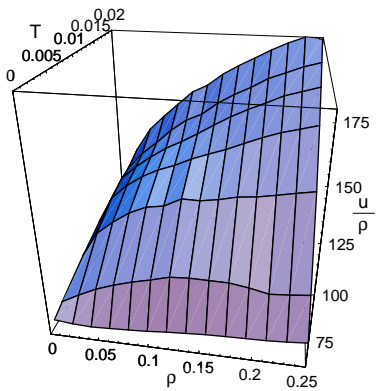}
\caption{The $\rho,~T,~\frac{u}{\rho}$ phase diagram. Some of the
contribution from finite-lattice effects is removed by using
Eq.~(\ref{FullScaled}) to find the internal-energy density. The
density $\rho$ is altered by changing the lattice spacing and keeping
the same number of Skyrmions in the system. The axes are labeled in
Skyrme units. (32~Skyrmions on an $80\times 80\times 80$ lattice.)}
\label{FTDRhoBetaEnergy}
\end{figure}

The internal energy per Skyrmion is plotted against the density $\rho$
and the temperature $T$ in Fig.~\ref{FTDRhoBetaEnergy}. The density
$\rho$ is determined by altering the lattice spacing $h$ in a
simulation with 32 Skyrmions on an $80\times 80\times 80$
lattice. Since some lattice-dependency is still present in the
internal energy of the system, the internal energies at different
densities may not be comparable, as they have been determined with
different lattice spacings. It is not possible to determine the states
of matter from Fig.~\ref{FTDRhoBetaEnergy}, and therefore alternative
methods will be used in the following sections.

\begin{figure}
\includegraphics[width=\graphwidth]{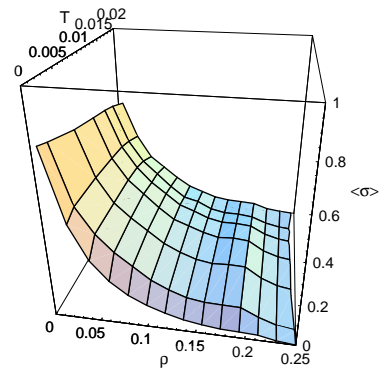}
\caption{The $\rho,~T,~\langle\sigma\rangle$ phase diagram. The kink
is a signature of the phase transition between the low-density and
high-density phases. The axes are labeled in Skyrme
units. (32~Skyrmions on an $80\times 80\times 80$ lattice with varying
lattice spacing.)}
\label{FTDRhoBetaChiral}
\end{figure}

The average sigma field is plotted against the density $\rho$ and the
temperature $T$ in Fig.~\ref{FTDRhoBetaChiral}. As already discussed
in the previous section for the baby-Skyrme model, a
chirally-symmetric phase exists at high densities,
$\langle\sigma\rangle=0$, and chiral symmetry is broken at low
densities, $\langle\sigma\rangle>0$. Although it is difficult to
determine the position of the phase transition between low-density and
high-density matter precisely, we believe it occurs where a kink can
be observed near $\rho=0.2$ for all temperatures plotted in
Fig.~\ref{FTDRhoBetaChiral}.

The pseudo-time-averaged baryon-density plots, with the help of
correlation functions, can be used to identify the different states of
matter. The solid, liquid and phase-coexistence states which are
observed can then be used to determine the phase portrait for the
Skyrme model. Since the Skyrme model is an approximation of QCD, the
phase portrait we create can be compared to QCD.

\begin{figure}
\includegraphics[width=\graphwidth]{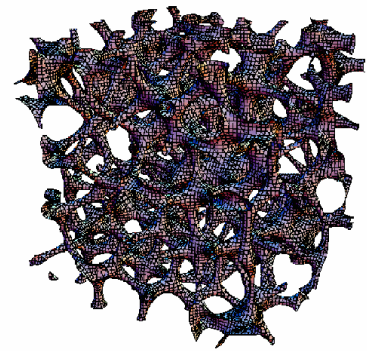}
\caption{A baryon-density contour plot, with a high baryon-density
contour, of a simulation at a high density
($\rho=0.244$~Skyrme-density units) and a finite temperature
($T=0.2$~Skyrme-energy units). An irregular, but crystalline,
structure can be seen. The mesh spacing is 0.08~Skyrme-length
units. (64~Skyrmions on an $80\times80\times80$ lattice with volume
$6.4^3$~cubic Skyrme-length units.)}
\label{HighFinite2}
\end{figure}

Examples of a contour plot from a pseudo-time-averaged baryon-density
distribution for a solid are shown in Fig.~\ref{HighFinite2} for a
high baryon-density contour. The crystal structure is irregular
because the simulation volume of $6.4^3$~cubic Skyrme-length units is too
small and therefore defects are introduced. We expected to observe a
hexagonal close-packed crystal of half-Skyrmions, but unfortunately no
regular crystal can be identified. In a low-density contour
plot~\cite{os:thesis}, we observe that the low-density regions are all
connected. This is in contrast to the disconnected circles of vacuum
observed for the baby-Skyrme model.

We have not shown an example of a contour plot of the
pseudo-time-averaged baryon-density distribution for a liquid.
These look rather misleading
when  only one contour is shown, as regions with
slightly higher- or lower-than-average baryon densities look like
Skyrmions. These structures which are due to small differences
in the pseudo-time-averaged baryon-number distribution
are actually artifacts of the simulation process.
When examining the results, we actually look at a variety of
contours to verify that we have a liquid, but this would be too
cumbersome to illustrate. A liquid can be distinguished from a solid
most easily by observing that a liquid does not have connected
low-density contours that look like tubes.

\begin{figure}
\includegraphics[width=\graphwidth]{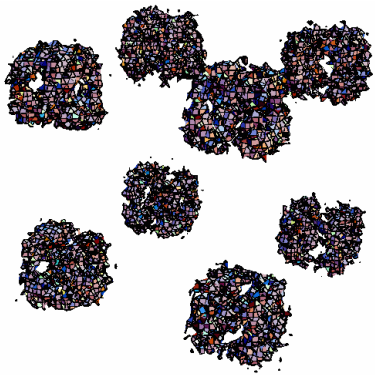}
\caption{A baryon-density contour plot of a simulation at a low
density ($\rho=0.122$~Skyrme-density units) and a finite temperature
($T=0.2$~Skyrme-energy units). Eight $B=4$ multi-Skyrmions can be seen
with temperature fluctuations and they would look like the $B=4$
multi-Skyrmions. The contour shown for each the $B=4$ multi-Skyrmion has an
approximate volume of 2.8~cubic Skyrme-length units. The mesh spacing
is 0.08~Skyrme-length units. (32~Skyrmions on an $80\times80\times80$
lattice with volume $6.4^3$~cubic Skyrme-length units.)}
\label{LowFinite}
\end{figure}

In the phase coexistence region, we observe multi-Skyrmions with large
regions of vacuum surrounding them. An example is shown by a snapshot
of a contour of the baryon-number distribution in
Fig.~\ref{LowFinite}. Here, eight $B=4$ multi-Skyrmions can be
seen. The $B=4$ multi-Skyrmions are very tightly
bound, like the alpha particles they model, which is why
multi-Skyrmions of lower winding number are not seen. The boundary
conditions also favor the $B=4$ multi-Skyrmions over other
multi-Skyrmions. Unfortunately, we are not able to run large enough
simulations to see if a mixture of other multi-Skyrmions can be
created. The regular spacing of the $B=4$ Skyrmions is most likely
also a consequence of a small number of Skyrmions in a limited
simulation volume.

\begin{figure}
\includegraphics[width=5cm]{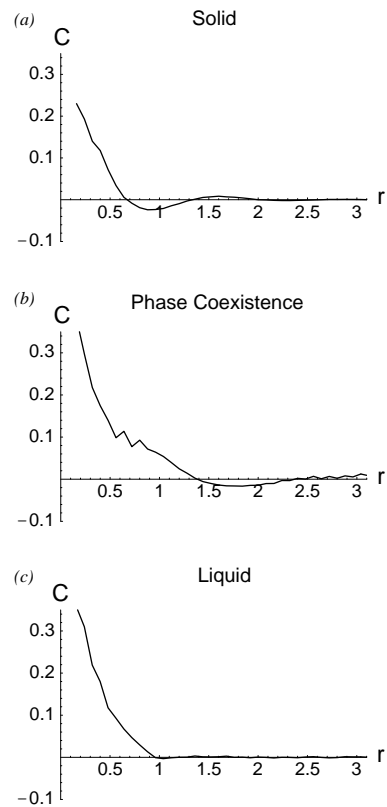}
\caption{The correlation functions for a solid, a state in phase
coexistence and a liquid, respectively. {\it (a):}~For a solid, the
correlations show long-range order at large distances. The correlation
function corresponds to the contour plots shown in
Fig.~\ref{HighFinite2}. {\it (b):}~For phase
coexistence, the self-correlations of the $B=4$ Skyrmions are clearly
visible for $r<1$. The correlation between the different $B=4$
multi-Skyrmions can be seen for $r>2.5$. {\it (c):}~The function falls
off faster than the function for the solid, but medium-range
interactions are still present and therefore it is not identified as a
gas. The correlation function corresponds to the contour plot shown in
Fig.~\ref{LowFinite}. ($r$ is in Skyrme-length units and $C$ is
dimensionless.)}
\label{CorrFull}
\end{figure}

\begin{figure}
\includegraphics[width=5cm]{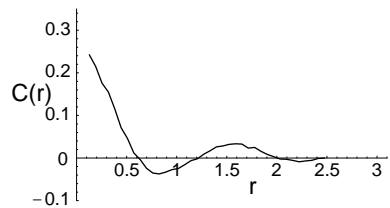}
\caption{The correlation functions for Castillejo {\it et al}
crystal, which is identified as a solid. ($r$ is in
Skyrme-length units and $C$ is dimensionless.)}
\label{FullCastCorr} 
\end{figure}

The same correlation function that is used for the baby-Skyrme model
can be applied to the Skyrme model, and has proven useful in the
analysis of the results.  The correlation functions for the solid,
liquid and phase coexistence states are given in
Fig.~\ref{CorrFull}. The correlation functions seem to fall off faster
for the Skyrme model than for the baby-Skyrme model, but for solids
and phase coexistence, a long-range order is clearly visible.  The
correlations for the putative solid must be compared to that of the
parametrised solution by Castillejo \emph{et al}, which consist of a
simple cubic lattice of half-Skyrmions~\cite{ca:dense}, is shown in
Fig.~\ref{FullCastCorr}. As one can see the decay of the correlation functions is very similar, if slightly less pronounced for the highly-defective
solid found in our simulations. For liquids,
there is also a visible medium-range order. The correlation of the
$B=4$ multi-Skyrmions with themselves and with other $B=4$
multi-Skyrmions are clearly visible (at short and long distances
respectively). Overall, the correlation functions for the Skyrme model show
the same type of behavior as for the baby-Skyrme model.

\begin{figure}
\includegraphics[width=\graphwidth]{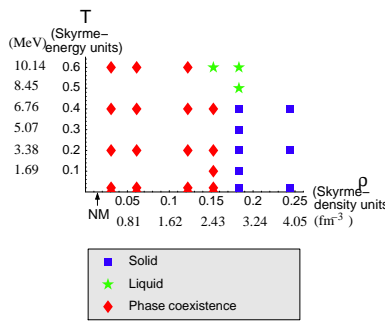}
\caption{The phase diagram of the Skyrme model including a pion-mass
term. The density of nuclear matter (NM) is marked at 0.16\,$\text{
fm}^{-3}$. Other models predict the chiral phase transition at
approximately 0.73\,$\text{fm}^{-3}$~\cite{ra:mapping}. All plotted
points are results obtained from simulations performed on a $80\times
80\times 80$ lattice with a lattice spacing of $h=0.08$ Skyrme-length
units.}
\label{FullPhaseDiagram} 
\end{figure} 

\begin{figure}
\includegraphics[width=5cm]{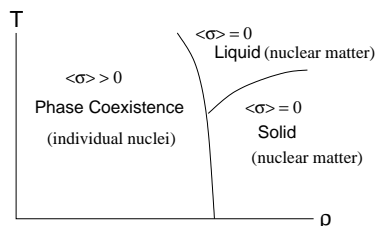}
\caption{A sketch of the phase diagram of the Skyrme model. The state
of matter and chiral symmetry (or symmetry breaking) is marked. The
corresponding phases of QCD are marked in parentheses.}
\label{FullPhaseDiagramSketch} 
\end{figure}

The final phase portrait for the Skyrme model is shown in
Fig.~\ref{FullPhaseDiagram}. The different states of matter were
distinguished from each other by examining the pseudo-time-averaged
baryon-density plots and the correlation functions. Classifying the
different states of matter in the vicinity where the different phases
meet is more difficult than for the baby-Skyrme model, because
long-range correlations are not as distinct. This means that the
assignments made at several points are somewhat
ambiguous. Nevertheless, the approximate location where different
phases meet has been identified. The chiral phase transition also
appears to separate the phase coexistence region with the liquid and
solid regions for this model at $\rho=0.18$\,Skyrme-length units for
the temperatures shown. Fig.~\ref{FullPhaseDiagramSketch} shows a
sketch of the phase diagram and the states of matter and whether there
is chiral symmetry is labeled. The solid and liquid phases are both
chirally symmetric and are identified with the quark-gluon plasma. The
coexistence phase is identified with nuclear matter. The phase
transition observed from chiral symmetry breaking and the phase
transition observed by looking at average baryon-density plots and
correlation functions are not necessarily signatures of the same phase
transition. If they are two different phase transitions, they are
close enough together so that we could not resolve the two. Two
different phase transitions may exist, because it may be possible to
create crystals that break chiral symmetry which are not in phase
coexistence with vacuum, which would be more identifiable with nuclear
matter.

\subsubsection{Conclusions and comparison with QCD}

The phase portrait of the Skyrme model including a pion-mass term
has been determined to a reasonable accuracy. The difficulties were
mainly caused because each simulation required a significant amount of
computing time. Furthermore, analysing three-dimensional structures
also requires more time and effort. Unfortunately, we did not have the
resources to extend the energy per baryon and chiral symmetry-breaking
phase diagrams to higher temperatures.

We will now compare our results from investigating the Skyrme model at
finite temperatures and finite densities to the phase portraits of QCD
obtained by other models and illustrated in
Sec.~\ref{Introduction}. The phase transition between nuclear matter
and deconfined quark matter occurs at $\rho=0.18$~Skyrme-density units
(2.9\,$\text{fm}^{-3}$). This phase transition, which has the
characteristic that chiral symmetry is broken, has been observed at
$\rho=0.045$~Skyrme-density units (0.73\,$\text{ fm}^{-3}$), which was
obtained from Ref.~\cite{be:color}. The phase transition occurs at
about 18 times the density of nuclear matter on earth of approximately
0.0099~Skyrme-density units (0.16\,$\text{
fm}^{-3}$)~\cite{wa:dense}. The lowest energy per baryon for QCD is
most likely at the normal density of nuclear matter, because any
excess energy would be radiated away. For the Skyrme model, the lowest
energy per Skyrmion occurs at $0.043$~Skyrme-density units
(0.70\,$\text{fm}^{-3}$). The lowest energy per baryon for the Skyrme
model and for QCD are not easily compared, because this value will
change significantly when the Skyrme model is quantised.

The lowest temperature where a liquid phase is observed is at
0.5~Skyrme-energy units (8.45\,MeV). Although this is approximately
the temperature of the critical point where nuclear matter can
continuously (i.e.\ without a phase transition) transform from a
liquid to a gas, we believe that the lowest temperature where a liquid
can be observed should be identified with the temperature of the
critical point where nuclear matter can transform to a quark-gluon
plasma continuously. These critical points  have been obtained from
Ref.~\cite{ra:mapping}. The reason we identify the lowest temperature
at which a liquid is observed with the temperature of the critical
point is because the high-density Skyrme matter, which behaves like a
solid, is identified with the quark-gluon plasma, as individual
Skyrmions are no longer distinct. The liquid phase may be the state
where high-density Skyrme matter (quark-gluon plasma) can continuously
be transformed to low-density Skyrme matter (nuclear matter). If this
is the case, then the predicted temperature of the critical point is
different by a factor of ten!

Alternatively, the liquid state for the Skyrme model may be identified
with the liquid state for the quark-gluon plasma, because the liquid
state observed for the Skyrme model is chirally symmetric. (Since
there are not many samples for the nuclear-Skyrme model, we observe
that this is also the case for the baby-Skyrme model.) If this is the
case, we have not been able to heat simulations enough to reach the
critical point. Furthermore, it is unclear whether the Skyrme model
shows any observable phenomena there. Possibly there is a liquid state
for all densities above this temperature.

The location of the phase transitions observed for the Skyrme model
will not correspond exactly with those of QCD, because the Skyrme
model is only a semi-classical approximation of QCD. However, it is
difficult to model the different phases of QCD by using only one
model. In future, the Skyrme model may be used to make statements
about the order of phase transitions in QCD. We were not able to
investigate this because we were only able to generate few samples
(which each required a considerable amount of computation time) and
even these samples are not statistically accurate enough to take
derivatives of the phase spaces that were generated. We have, however,
shown that field theories containing solitons have an interesting
phase portrait.

\section{Conclusions and Outlook}\label{sec:conclusions}

The phase portraits for the baby-Skyrme and nuclear-Skyrme models were
created accurately using the canonical approach. We have shown that
the low-density and high-density phases can be distinguished through
chiral symmetry breaking. Furthermore, pseudo-time-averaged
baryon-density plots and correlation functions can be used to identify
the state of matter at a given temperature and density. The
baby-Skyrme model has a rich phase diagram. Although the model does
not have a physical interpretation, we believe that similar phases
exist for Skyrmions in the quantum Hall effect. The phase diagram for
the nuclear-Skyrme model was compared to other models of the QCD phase
diagram. We were able to identify the phase where chiral symmetry is
broken with hadronic matter and the chirally symmetric phase with the
quark-gluon plasma. The location of the phase transition is at a
higher density than predicted by other models. It remains unclear
whether the boundary of the phase coexistence region with the liquid
and solid regions occurs on the same line as the chiral phase
transition.

The grand-canonical approach could not successfully be used to
generate accurate phase portraits for the baby-Skyrme and
nuclear-Skyrme models, but it is useful in determining the natural
crystal structure for the baby-Skyrme model. The open boundaries in
this approach allow crystals to form without defects. The
grand-canonical approach is already being applied to QHE
simulations~\cite{we:QHE}. Unfortunately, this method was unsuccessful
for the nuclear-Skyrme model, partly because we did not have the
necessary computer resources. We also were not able to run simulations
for a large number of different parameters. We believe that it is
possible to determine the natural crystal structure for the
nuclear-Skyrme model, and therefore propose that this be attempted
again. We suggest to remove the pion-mass term in such calculations,
because the minimal energy per baryon occurs in the high-density phase
rather than the low-density phase (as is the case when a pion-mass
term is included). When using the grand-canonical approach, it may be
effective to use an approximation to the derivative that results in
overestimating the baryon number, rather than underestimating like we
have done.

The thermodynamic investigations can also be applied for the Skyrme
model with a sixth-order term and for the $\omega$-stabilized Skyrme
model. The higher-order terms become dominant at high densities, and
may therefore be better models for describing such phases.

The nuclear-Skyrme model is a semi-classical approximation to QCD in
the $N_c\rightarrow\infty$ limit. This approximation is not
particularly accurate, but it means that a single model can be used to
describe matter at both low densities and high densities. We were not
able to investigate the order of the phase transitions, and since this
may be relevant to QCD, it would also be interesting subject for
future research.

Thus, we have derived methods to investigate models with soliton
solutions for both zero and finite temperatures. The methods we
developed can easily be applied to a wide range of similar models.

\acknowledgments
OS has been supported by an EPSRC studentship;
NRW acknowledges support from the EPSRC through research grants GR/L22331
and GR/N15672.


\end{document}